# MX precipitate behavior in an irradiated advanced Fe-9Cr steel: Helium effects on phase stability

T.M. Kelsy Green[a*], Tim Graening[b], Weicheng Zhong[b], Ying Yang[b], and Kevin G. Field[a]

[a]University of Michigan-Ann Arbor, currently at Antares Nuclear Industries
[b]Oak Ridge National Laboratory
[a]University of Michigan-Ann Arbor

*Corresponding Author: kelsy@antaresindustries.com

## Graphical Abstract

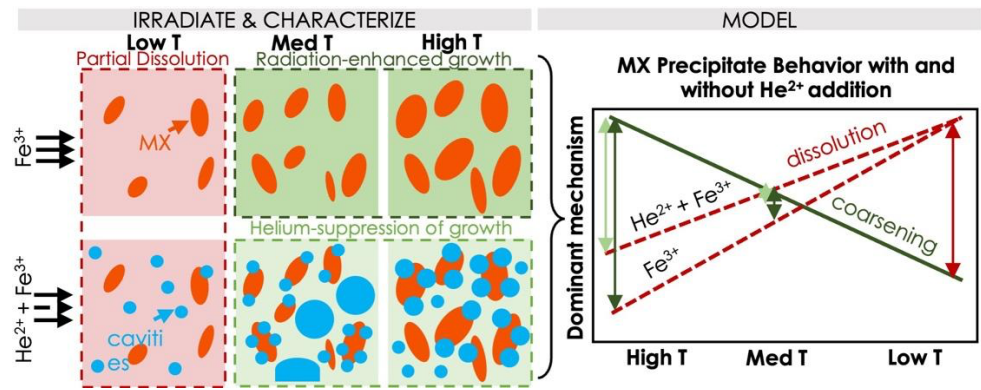

## Abstract

Precipitates are main microstructural features to provide high temperature creep strength and radiation resistance in structural materials for fusion energy systems. However, the mechanisms of precipitate stability under irradiation in candidate structural materials for fusion first-wall and blanket components are poorly understood. In particular, the dual effects of helium transmutation and irradiation-induced damage on precipitate evolution have not been systematically studied in candidate materials, the leading of which are Fe-9Cr reduced activation ferritic/martensitic (RAFM) alloys. To fill this knowledge gap, a fundamental understanding of the single and combined interactions of helium (0-25 appm He/dpa), temperature (300-600°C), and atomic displacements (15-100 dpa) on the behavior of MX (M=metal, X=C and/or N) precipitates in an advanced Fe-9Cr RAFM alloy were studied through the use of dual ion irradiation experiments. It was found that helium suppressed the diffusion-mediated mechanisms of precipitate stability (*i.e.* radiation-enhanced growth) at elevated temperatures and intermediate damage levels but had no effect on precipitate dissolution in the high dose conditions (≥50 dpa). A precipitate stability model was used to rationalize the impacts of helium on ballistic dissolution and radiation-enhanced diffusion which are key contributors to overall precipitate stability. This is the second paper in a series of three to provide a systematic evaluation of MX precipitate behavior in RAFM steels under various fusion-relevant ion irradiation conditions.

## 1.1 Introduction

Fusion first-wall and blanket components are the underlying structures that support the plasma-facing components. They support heat extraction for energy production and tritium breeding for sustained operation. The effects of high energy neutrons produced in the deuterium-tritium (D-T) plasma reaction in a fusion reactor core on first-wall and blanket components are two-fold: 1) the neutrons and subsequent displaced atoms cause damage to the lattice of the materials and 2) the neutrons cause helium transmutation if the any atoms within the material composition has a (n,$\alpha$) cross-section [1]. Both radiation damage and helium-induced degradation will fundamentally alter the microstructure of the materials, ultimately leading to changes in macroscopic properties [2]. Of particular scientific interest are the various effects of the transmuted helium on structural materials: increased cavity swelling (leading to dimensional instability), changes in solute point defect segregation, altered secondary phase evolution, and modified dislocation structures [3-9].

Due to superior mechanical properties and reduced waste concerns from lower activation under neutron irradiation, body centered cubic (BCC) reduced activation FM (RAFM) steels are currently the foremost alloys being considered for use as fusion structural materials [10]. RAFM steels are favored for fusion applications due to their hierarchical grain structure, precipitation structures, and high dislocation densities [11]. RAFM steels are Fe-based alloys with ~7-9 wt.% Cr and ~1-3 wt.% of alloying elements such as W, V, N, C, Mn, and Si. The normalized and tempered state of RAFM steels contains MX precipitates (M=metal such as Ti, Ta, V, or Nb; X=C and/or N) on and within grain boundaries [12]. Originally developed to increase high temperature creep strength by pinning dislocations and grain boundaries [13, 14], abundant MX precipitation also has the potential to increase radiation resistance, measured by sink strength, of RAFM steels. In general, higher sink strength materials ($\geq 10^{16}$ m$^{-2}$) have been theorized to correlate to superior lattice damage and helium resistance, in terms of both longer swelling incubation times and lower steady-state swelling rates, as compared to lower sink strength materials [15-19]. Hence, engineering small precipitates (<10 nm in diameter) in high number densities (>$10^{21-22}$ m$^{-3}$) is an important alloy design strategy to mitigate the negative effects of damage and helium transmutation on RAFM steels.

The most recent RAFM steels developed in the United States are the family of Castable Nanostructured Alloys (CNAs). CNAs aim to increase MX precipitation as compared to previous generations of Fe-9Cr RAFM steels to improve sink strength and swelling resistance. The variant of CNAs called CNA9 was able to achieve a number density of $(2.3\pm0.3)\times 10^{21}$ m$^{-3}$ of MX-TiC precipitates, which is about 1-2 orders of magnitude greater than MX precipitates found in other Fe-9Cr steels [20, 21], with an average equivalent diameter of 7.8±0.3 nm [20]. CNA9 contains no $M_{23}C_6$ precipitates [22]. In order for the MX precipitates to provide beneficial properties over the lifetime of a reactor, their stability under irradiation to high doses ($\geq$100 dpa) needs to be systematically studied in fusion-like conditions. Though the relationship between helium and irradiation effects on MX precipitates has been examined in austenitic steels [4, 5, 8, 23-29], no fundamental studies in CNAs or other RAFM steels have been conducted to understand if the presence of helium under irradiation alters the stability of MX precipitates which have been engineered to provide improved mechanical properties and helium resistance. Knowledge gaps remain on the synergistic roles of radiation damage and helium transmutation on the performance of CNA9 steel (and RAFM steels in general) [2, 30, 31].

This work examines the helium-impacted microstructural behavior in CNA9 to understand the co-evolution of helium and MX precipitates in RAFM steels using single and dual beam ion irradiations to mimic fusion first-wall and blanket operation conditions. Previous work found that MX precipitate behavior in CNA9 under single beam irradiation was temperature-dominated (dissolution below 400ºC, coarsening above 500ºC) at intermediate damage level (15 dpa) but that all precipitates dissolved by 50 dpa independent of the temperatures tested [22]. The current work will expand upon the previous work by investigating the effects of helium co-implantation at the same temperature and damage levels using high fidelity dual ion (Fe + He) irradiation experiments and advanced electron microscopy. Note, the implications of cavity swelling are part of the authors third and final component of this series. Experimental results will be compared to results from a precipitate stability model [32].

## 1.2 Methods
### 1.2.1 Material

A laboratory-scale advanced engineering alloy called Castable Nanostructured Alloy #9 (CNA9) was used in this work. Ref. [22] provides extended details on CNA9 and the broader family of CNA Steels. The composition of CNA9 in weight percent is provided in Table 1. This composition was developed to promote a high number density of MX-TiC precipitates via a tailored thermomechanical treatment of normalization at 1050°C for 1 h in Ar atmosphere followed by hot-rolling at 1050°C and tempering at 750°C for 30 min [33]. The final precipitate population included large, spherical TiC precipitates with a diameter range of ~50-100 nm and small MX-TiC precipitates with an equivalent diameter of 7.9±0.3 nm, which are the primary focus of this work [22, 34]. The smaller scale MX-TiC type precipitates existed both inter- and intra-lath in the matrix. All samples for this study were produced by electrical discharge machining (EDM) from the laboratory scale ingot followed by standard metallographic procedures with the final polished state achieved via electropolishing using a -45°C cooled solution of 10% perchloric acid solution and 90% methanol solution.

Table 1 Chemical compositions (wt%) of CNA9 provided by Dirats Laboratories.

| Element | CNA9 (wt.%) | Element | CNA9 (wt.%) |
|---|---|---|---|
| Fe | 89.27±0.02 | Co | <0.005±0.0005 |
| Cr | 8.688±0.08688 | Cu | <0.002±0.0002 |
| W | 1.026±0.0126 | Mo | 0.004±0.0004 |
| Mn | 0.516±0.00516 | Nb | <0.002±0.0004 |
| Si | 0.141±0.00141 | Ni | <0.007±0.0007 |
| Ta | 0.090±0.0009 | P | 0.004±0.0004 |
| Ti | 0.141±0.00141 | Zr | <0.002±0.0002 |
| V | 0.049±0.00049 | S | 0.002±0.0002 |
| C | 0.049±0.00049 | O | 0.0012±0.00012 |
| Al | <0.002±0.0002 | N | 0.0013±0.00013 |
| B | <0.0005±0.00005 | | |

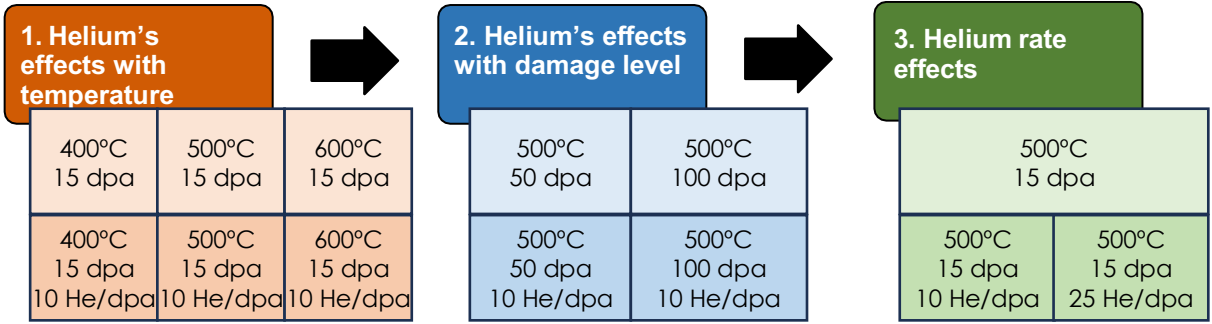

Figure 1 Graphical representation of the irradiation experiments conducted to understand the effects of helium co- implantation on MX phase stability during ion irradiations. All irradiations were conducted with a damage rate of $7\times10^{-4}$ dpa/s.

### 1.2.2 Ion irradiation Experiments

The ion irradiation experiments were designed to test the single and combined effects of temperature, damage level, and helium co-implantation. Refer to Ref. [35] for a detailed discussion on the results of the single beam, self-ion only irradiations. This work will focus on the effects of helium co-implantation as a means of simulating the (n,α) reactions under neutron irradiation by discussing three series of experiments: the temperature series, the damage level series, and the helium co-implantation series (Figure 1). The irradiation parameters chosen are relevant to fusion first-wall and blanket environments while also providing a granular look at the fundamental behavior of precipitates, thus allowing for both engineering and scientific insights. The target matrix damage rate used in each irradiation was $7\times10^{-4}$ dpa/s [22] and the target matrix damage levels were 15, 50, 100 dpa. The target temperatures used were 400, 500, and 600ºC. This temperature range encompasses the known peak swelling temperature for FM steels (~420-500ºC) [36-38]. The helium co-implantation rates in the nominal damage region used in the dual beam irradiations were 10 and 25 appm He/dpa, as these mimic first-wall helium production [39, 40].

The dual beam configuration at the University of Michigan-Ann Arbor's Michigan Ion Beam Laboratory (MIBL) was utilized for all experiments. Damage was induced from a flux of 9 MeV $Fe^{3+}$ ions impinging perpendicular to the sample surfaces while the 3.42 MeV $He^{2+}$ ions were implanted at an angle of 60° from the surface of the samples. The Stopping and Range of Ions in Matter (SRIM 2013) [41] was utilized to calculate the damage of the $Fe^{3+}$ ions and the helium implantation profile of the $He^{2+}$ ions. The SRIM damage calculations were conducted using "quick" Kinchin-Pease (KP) mode [38] (Figure 2). The rate of helium co-implantation was controlled by a rotating aluminum foil degrader which was run with a LabView™ script during irradiations [42, 43]. The degrader system maintains a constant helium/dpa ratio in the nominal damage region, as seen in Figure 2. The nominal damage region was between 1,100-1,300 nm beneath the surfaces of the bulk irradiated specimens where the target matrix damage level, damage rate, and helium co-implantation were reached. During the irradiations, temperature, pressure, beam current, and beam profiles were recorded and displayed on computers using custom built LabView™ programs. Temperature was controlled to within ±10ºC of the target value during experiments with the use of an infrared thermal pyrometer and thermocouple readings. The beam currents were controlled to within ±10% of the desired currents.

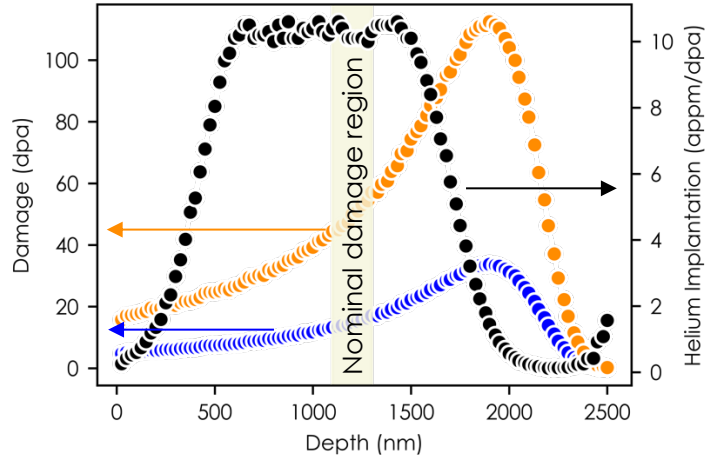

Figure 2 The damage profiles for the 15 (blue line) and 50 dpa (orange line) irradiations as well as the helium implantation rate in appm/dpa (black line). The target damage and helium rate levels were reached in the nominal damage region 1,100-1,300 nm beneath the surface.

### 1.2.3 Sample preparation and characterization

Refer to Ref. [22] for a detailed explanation of sample preparation and characterization. Briefly, standard focused ion beam (FIB) milling practices were used to extract cross-sectional slices of irradiated CNA9 bulk material, of which encompassed both the nominal damage region and the region past the $Fe^{3+}$ ion implantation depth where only thermal effects were present. To tabulate MX-TiC precipitate diameter ($d$), number density ($\rho$), and volume fraction ($f$) as a function of irradiation parameters, Ti compositional mapping was used on a scanning transmission electron microscope (STEM) equipped with Energy Dispersive X-ray Spectroscopy (EDS) capability. The major axis, $a$, and minor axis, $b$, of each MX-TiC precipitate present in the STEM-EDS maps were measured in a freely available image processing suite called FIJI, or ImageJ [36]. The equivalent diameter was then calculated as $d = \sqrt{a \times b}$. The size distributions of precipitates are visualized using width-scaled violin plots generated using the Python 3.7 Seaborn violin plot function [44], which uses a Gaussian kernel density estimation.

## 1.3 Results and Discussion

### 1.3.1 Irradiation and helium effects on MX-TiC precipitation

*1.3.1.1 Helium co-implantation effects at intermediate damage level (15 dpa) with variable temperature (400, 500, and 600ºC)*

The prior analysis conducted in the previously presented work established the baseline control condition from which to compare to the irradiated MX precipitation at 15 dpa [22]. The number density ($\rho_{CTRL}$) of precipitates in the control condition was $(2.7\pm0.3)\times10^{21}$ m$^{-3}$, the equivalent diameter ($d_{CTRL}$) was $7.9\pm0.3$ nm, and the volume fraction ($f_{CTRL}$) was $(9.0\pm1.4)\times10^{-4}$. The MX-TiC precipitates are shown in Figure 3a and b, and the precipitate size distribution of the control condition is shown in Figure 3c. The dashed lines on the violin plots represent the 25% and 75% interquartile lines. The mean equivalent diameter with error bars representing the standard error is also shown on the size distribution. The response of precipitates to irradiation will be quantified by comparing the irradiated precipitate size distribution and statistics to those from the control condition: $\rho_{IRR}/\rho_{CTRL}$, $d_{IRR}/d_{CTRL}$, and $f_{IRR}/f_{CTRL}$, where *'IRR'* refers to irradiated. To quantify if the changes in statistics were statistically significant with irradiation, these ratios will be compared to values tabulated in Table 2. Refer to Ref. [22] for a more detailed explanation of the analysis of the control condition. Refer to Supplemental A for all data on precipitate statistics in each dual ion irradiated condition (*i.e.*, number of EDS maps taken, number of precipitates counted, etc.).

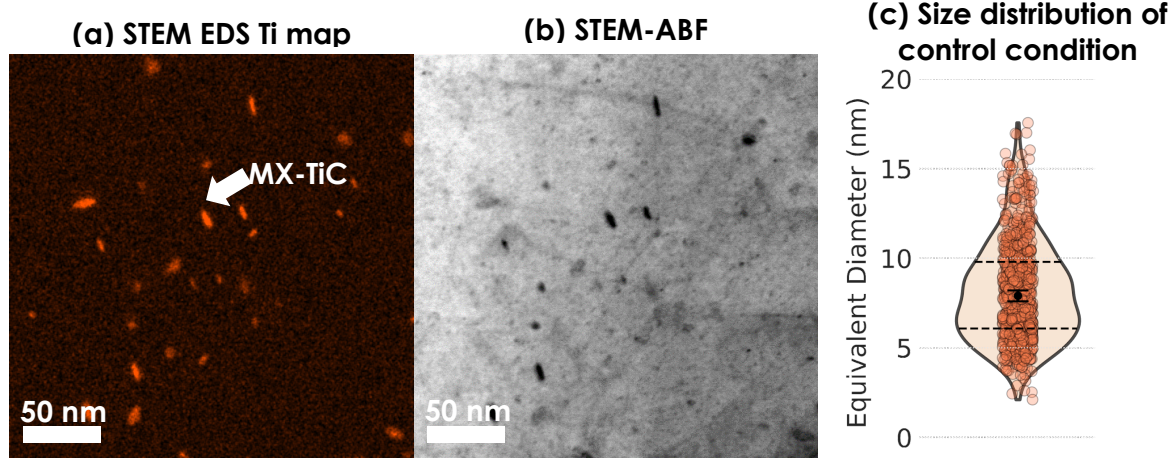

Figure 3 Example (a) STEM-EDS Ti map with corresponding (b) STEM-ABF micrograph of the control condition. (c) Size distribution of the MX precipitates present in the control condition.

Table 2 Tabulated values that represent the range of statistical significance for ratio calculations between irradiated and control conditions. Reproduced from Ref. [22]. Refer to Ref. [22] for a more detailed explanation.

| Ratio | Significant if less than | Significant if greater than |
|---|---|---|
| $\rho_{IRR}/\rho_{CTRL}$ | 0.6 | 1.7 |
| $d_{IRR}/d_{CTRL}$ | 0.8 | 1.3 |
| $f_{IRR}/f_{CTRL}$ | 0.4 | 2.5 |

The single beam irradiations were conducted at 400, 500, and 600°C to 15 dpa with no helium co-implantation. The MX-TiC precipitates displayed temperature-dominated behavior at this intermediate damage level during single beam irradiations: the partial dissolution of precipitates occurred at the condition irradiated to 400°C at 15 dpa with no precipitate size changes and, contrastingly, the radiation-enhanced coarsening of MX-TiC precipitates was observed at 500 and 600°C, most notably at 600°C.

The effects of helium co-implantation on precipitate behavior are shown in Figure 4. Figure 4 shows example STEM-EDS Ti maps for the dual ion irradiated specimens with the corresponding size distribution of precipitates for each condition in the temperature series examined. The statistical precipitate response as compared to the control condition is displayed in the table below the size distribution plots, where a ratio less than 1 means that variable decreased with irradiation and a ratio greater than 1 means that ratio increased with irradiation. The direct comparisons of the single and dual beam precipitate size distributions and statistics for the samples irradiated to 15 dpa are shown in Figure 5, where the single beam precipitate size distributions are shown on the left-hand sides of the split violin plots and the dual beam size distributions on the right-hand sides. The statistical responses in Figure 5 compare the dual to the single beam precipitate statistics: $\rho_{DI}/\rho_{SI}$, $d_{DI}/d_{SI}$, and $f_{DI}/f_{SI}$.

It can be seen from Figure 4 and Figure 5 that the 400ºC dual beam condition irradiated to 15 dpa displayed the same behavior as the corresponding single beam irradiation: the statistically significant partial dissolution of precipitates determined from the change in number density but no precipitate size changes as compared to the control condition.

By contrast, the 600ºC dual beam condition irradiated to 15 dpa still underwent statistically significant radiation-enhanced coarsening seen in the single beam condition (Figure 4) but the degree of coarsening was suppressed with the presence of helium in comparison to the corresponding single beam condition (Figure 5). This observation is ascertained by the suppression of the dual beam precipitate distribution where the maximum shifted downward from ~35 nm to ~28 nm, the 75% interquartile value from ~20 to ~15, and the mean value from ~15 nm to ~11 nm (Figure 5).

The same mechanism of the suppression of radiation-enhanced coarsening at 600ºC is also observed in the 500ºC dual beam condition, except the coarsening is fully suppressed at 500ºC (Figure 5) and the size distribution is now nearly equal to the control size distribution (Figure 4). Though the dual beam size distributions at 400 and 500ºC are of similar shape, the 500ºC condition did not undergo significant partial dissolution, suggesting helium co-implantation was able to suppress coarsening but not impact dissolution at this dose condition. Thus, helium co-implantation had no observable effect on the MX-TiC precipitates at 400ºC to intermediate damage level (15 dpa) but actively suppressed radiation-enhanced coarsening at 500 and 600ºC.

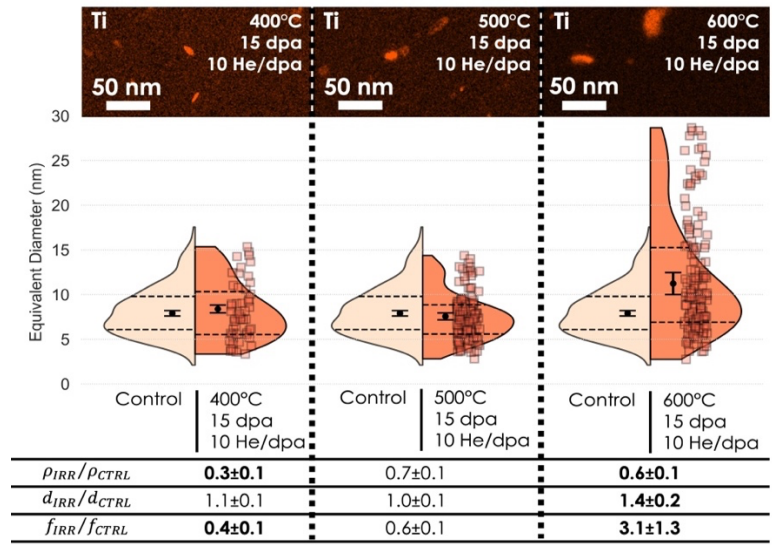

Figure 4 STEM-EDS micrographs for the control and each irradiated conditions in the dual beam temperature series, along with the corresponding split violin plots and ratios of number density ($\rho$), average equivalent diameter (d), and volume fraction (f). Statistically significant values are bolded in the table.

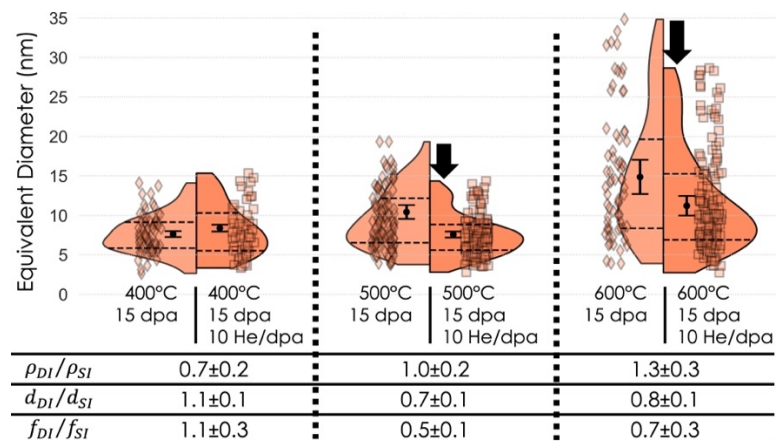

Figure 5 The comparison in responses between the single and dual beam conditions at 400, 500, and 600ºC. The ratios now show the change in parameters from the dual to the single beam conditions. Diamonds represent precipitates from single beam irradiations and squares from dual beam irradiations. The black arrows indicate the suppression of the dual beam size distributions.

*1.3.1.2 Helium co-implantation effects at variable high damage level (50, 100 dpa) with constant temperature (500ºC)*

The effect of helium co-implantation on precipitate behavior at higher damage levels will now be assessed. To study this, four irradiations were completed in the damage level series at 500ºC to 50 and 100 dpa without helium co-implantation and with 10 appm He/dpa. The temperature of 500ºC was chosen for further study because precipitates showed helium-induced stability when irradiated at this temperature at 15 dpa. In addition, 500ºC was nearest to the hypothesized peak swelling temperature out of the three temperatures assessed, and swelling at higher damage levels is integral to the findings reported in the third paper in this series.

As can be seen in Figure 6a and b, the MX-TiC precipitates dissolved by 50 dpa and remained dissolved at 100 dpa with no helium co-implantation. Likewise, the precipitates dissolved at 50 and 100 dpa with the addition 10 appm He/dpa (Figure 6c and d). Thus, helium co-implantation did not alter the complete precipitate dissolution at the damage levels studied. Importantly, helium may have lengthened the onset of dissolution at a damage level between 15 and 50 dpa, but obtaining this value does not have engineering-relevant consequences as damage levels less than 50 dpa represent a short span in the lifetime of a fusion reactor.

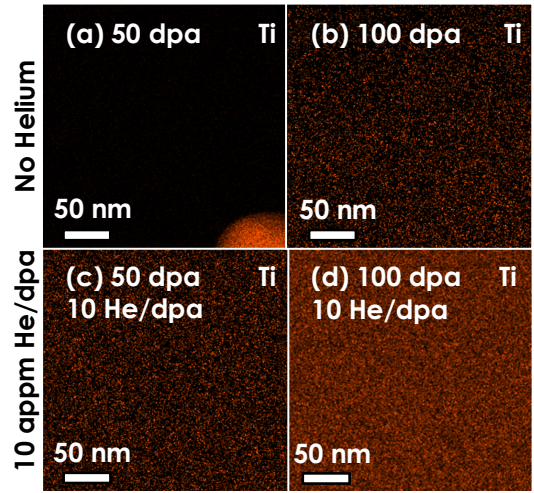

Figure 6 STEM-EDS micrographs of Ti for single beam conditions irradiated to (a) 50 and (b) 100 dpa. STEM-EDS micrographs of Ti for dual beam conditions irradiated to (c) 50 and (d) 100 dpa with 10 appm He/dpa. All conditions were irradiated at 500°C and rate $7\times10^{-4}$ dpa/s.

*1.3.1.3 Helium co-implantation effects at variable helium rates and constant temperature*

The irradiation temperature of 500°C was shown to be a pivotal point in phase stability with 10 appm He/dpa co-implantation at 15 dpa (*i.e.* helium co-injection fully suppressed radiation-enhanced coarsening present in the single beam condition) (Figure 5). To test the effects of a greater helium implantation rate on MX phase stability, a helium co-injection of 25 appm He/dpa at 500°C to 15 dpa was chosen. The results are shown in Figure 7, where the precipitate response irradiated at 500°C to 15 dpa with no helium and with 10 and 25 appm He/dpa is shown.

The size distribution and statistics of the MX-TiC precipitation in the 25 appm He/dpa condition were nearly identical to the precipitation present in the 10 appm He/dpa condition. The larger precipitate sizes (≥15 nm) present in the single beam condition at 500°C were suppressed in both dual beam conditions. It is hypothesized, in line with previous research on the effects of complex microstructures on macroscopic properties [18], that the high sink strength of the CNA9 matrix (~$10^{14}$-$10^{15}$ m$^{-2}$) was able to effectively accommodate the total helium concentrations of 150 and 375 appm He in the dual beam irradiations without entering different swelling regimes.

Basically, CNA9 remained in the incubation and transient regimes in both irradiations. Perhaps a greater helium rate would force the onset of the linear steady-state swelling regime by 15 dpa. However, 25 appm He/dpa was the upper limit achievable at MIBL and is relevant to the helium generation range for structural steels for fusion reactors.

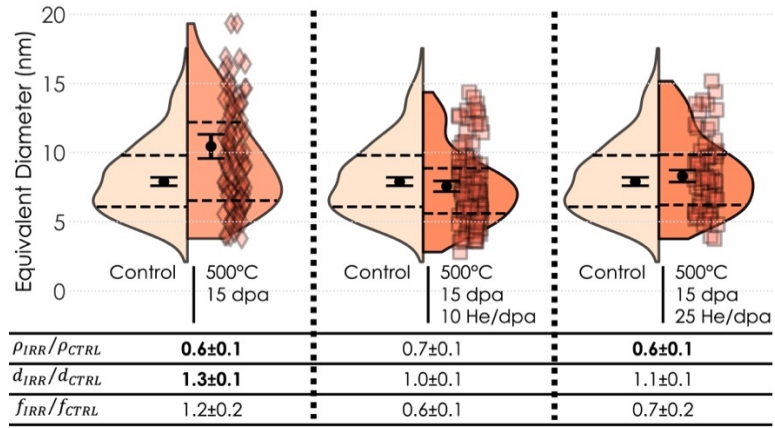

Figure 7 Statistics for the helium co-injection series at 500°C (0, 10, 25 appm He/dpa) as compared to the control specimen.

### *1.3.2 Application of the recoil resolution model of precipitate stability for MX-TiC precipitates in irradiated CNA9*

As described in the previous sections, radiation recoil was dominant at 400ºC to 15 dpa with and without helium whereas growth via radiation-enhanced diffusion was dominant at the 500 and 600ºC single beam conditions to 15 dpa. Helium co-implantation acted to suppress the radiation-enhanced coarsening at 500 and 600ºC. Thus, helium appears to affect the diffusion-mediated mechanisms of precipitate stability. To understand mechanistically how helium's effects on diffusion may alter precipitate stability, the recoil resolution model of precipitate stability under irradiation originally proposed by Frost and Russell can be used [32].

Precipitate stability under irradiation is typically described as a balance between resolution via radiation recoil (also known as ballistic dissolution) as well as growth and coarsening via radiation-enhanced diffusion. Resolution by radiation recoil hastens dissolution via radiation damage causing solute atoms in a precipitate to recoil into the matrix. Diffusion would hasten ripening processes like Ostwald ripening and also coarsening mechanisms if free and excess Ti & C still exist within the matrix. The recoil resolution model is an expansion of the work by Nelson and considers the solute deposition rate into the matrix to be a function of proximity to the precipitate, which was not previously included in the original Nelson's model [30, 31]. The model determines the dominant mode of precipitate stability between the competing mechanisms of recoil resolution and diffusion by calculating the matrix concentration profile of the rate-limiting species for precipitation near the precipitate-matrix interface. For MX-TiC precipitates in CNA9, the precipitate rate-limiting species was determined to be Ti (see Supplemental B). The Ti solute concentration in the matrix near the precipitate-matrix interface ($c_{max}(r_p + R)$) is derived by solving the steady-state Ti concentration profile from the diffusion equation, modified by the rate of recoil resolution:

$$c_{max}(r_p + R) = c_e + \frac{SR^2}{12D}\left(1 - \frac{R}{4r_p}\right) \qquad \text{Eq. 2}$$

where $R$ is the recoil distance, $S$ is the recoil rate (in dpa/s), $D$ is the diffusion of the rate-limiting species (Ti in this analysis), $r_p$ is the mean precipitate radii determined experimentally at 15 dpa, and $c_e$ is the thermal concentration (in atom fraction) of Ti calculated from Thermo-Calc. The recoil distance $R$ refers to the average weighted distance that a recoiled solute atom travels outside of a precipitate after a collision. This value was determined with SRIM to be 0.7 nm [22]. Coarsening is assumed to be the dominant mode of precipitate stability if the Ti matrix concentration from thermal equilibrium is greater than the concentration from recoil resolution, and vice versa. Importantly, the model does not consider the effects of solutes recoiling from the matrix into the precipitate, the model assumes a unimodal size distribution of pre-existing particles, and the model assumes equal and regular spacing between pre-existing particles – all of which may affect the fidelity of the model results.

The ability to decipher the effects of helium co-injection on precipitate behavior stems from how Ti solute diffusion ($D$) in Eq. 2 is calculated. Helium atoms (both interstitial and substitutional) bind to matrix atoms and suppress their diffusion, and the level of binding of helium to the matrix atoms ($E_b^{He}$) depends on the type of helium atom and matrix atom, temperature of the system, and the presence of point defects (such as vacancies). Relevant to CNA9, literature has shown that helium is attracted to Ti solutes in BCC Fe systems [6-8, 45]. To quantitatively assess how helium affects the Ti solute diffusion and hence the MX-TiC precipitate stability, the following equations were used. First, the thermal diffusion was calculated [3]:

$$D_{thermal}^{Ti} = D_{v,thermal}^{Ti} C_v^0 + D_{i,thermal}^{Ti} C_i^0 \qquad \text{Eq. 3}$$

$$D_{(v,i),thermal}^{Ti} = \alpha a^2 v \exp\left(\frac{S_m^{v,i}}{k}\right) \exp\left(\frac{-E_m^{v,i}}{kT}\right) \qquad \text{Eq. 4}$$

$$C_{v,i}^0 = \exp\left(\frac{S_f^{v,i}}{k}\right) \exp\left(\frac{-E_f^{v,i}}{kT}\right) \qquad \text{Eq. 5}$$

The diffusion of Ti under single beam irradiation was then calculated from the equation for the radiation-enhanced diffusion of a solute [44]:

$$D_{RED}^{Ti} = D_{v,thermal}^{Ti} \frac{C_v^{irr}}{C_v^0} + D_{i,thermal}^{Ti} \frac{C_i^{irr}}{C_i^0} \qquad \text{Eq. 6}$$

The diffusion of Ti in the presence of helium can then be calculated from:

$$D_{RED,He}^{Ti} = \frac{D_{RED}^{Ti}}{1 + c_e \exp\left(\frac{E_b^{He}}{kT}\right)} \text{ [45]} \qquad \text{Eq. 7}$$

In the above equations, $C_v^0$ is the thermal equilibrium concentration of vacancies, $C_i^0$ is the thermal equilibrium concentration of interstitials, $C_v^{irr}$ is the concentration of vacancies under irradiation, and $C_i^{irr}$ is the concentration of interstitials under irradiation. Refer to Supplemental C for how these values were calculated. The variables used to calculate these concentrations are given in Table 3.

Table 3 Variables and their values used in the diffusion calculations.

| Variable [3] | Symbol | Value |
|---|---|---|
| Jump frequency | $\alpha$ | Vacancies: $1.6 \times 10^{13}$ Hz<br>Interstitials: $2.9 \times 10^{12}$ Hz |
| Lattice parameter of the matrix | $a$ | 2.96 Å |
| Interstitial formation entropy | $S_f^v$ | 2.17 ev/K |
| Vacancy formation entropy | $S_m^i$ | 0 eV/K |
| Interstitial formation energy | $E_f^v$ | 1.6 eV |
| Vacancy formation energy | $E_f^i$ | 5 eV |
| Interstitial migration energy | $E_m^v$ | 0.62 eV |
| Vacancy migration energy | $E_m^i$ | 0.35 eV |
| Binding energy of the first nearest-neighbor substitutional helium with a Ti solute in BCC Fe [6] | $E_b^{He}$ | 0.34 eV |

Thus, the radiation-enhanced diffusion of Ti ($D_{RED}^{Ti}$, Eq. 6) was used for Ti solute diffusion in single beam irradiations and the helium-affected radiation-enhanced diffusion of Ti ($D_{RED,He}^{Ti}$, Eq. 7) was used for Ti solute diffusion in dual beam irradiations. In this way, the values of $c_{max}(r_p + R)$ under thermal ($c_e$), single beam irradiation ($\frac{SR^2}{12D_{RED}^{Ti}}\left(1 - \frac{R}{4r_p}\right)$), and dual beam irradiation ($\frac{SR^2}{12D_{RED,He}^{Ti}}\left(1 - \frac{R}{4r_p}\right)$) can be calculated and the effects of helium co-injection on precipitate behavior understood. As the dual beam conditions at 500°C with 10 and 25 appm He/dpa showed no difference in precipitate behavior, only the 10 appm He/dpa conditions will be assessed.

First, the diffusion values for thermal, single beam, and dual beam conditions were calculated and plotted for 400, 500, and 600°C in Figure 8. It can be observed that radiation-enhanced diffusion, $D_{RED}^{Ti}$, which is operational under single beam irradiation, is dominant up to ~550°C due to the greater concentration of point defects under irradiation. The concentration of vacancies increases with increasing temperature, leading to nearly similar Ti diffusion values by 600°C under thermal and irradiation conditions. $D_{RED,He}^{Ti}$ is smaller in magnitude than $D_{RED}^{Ti}$, showing the effect of helium suppression on diffusion.

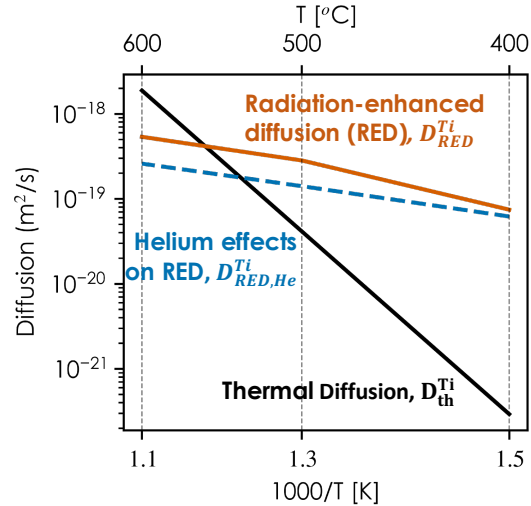

Figure 8 Thermal diffusion ($D_{th}^{Ti}$, black line), radiation-enhanced diffusion in single beam conditions ($D_{RED}^{Ti}$, orange line), and helium-suppressed diffusion in dual beam conditions ($D_{RED,eff}^{Ti-He}$ and $D_{RED,eff}^{Ti-He-v}$, blue lines).

Using this understanding of diffusion, the precipitate behavior was plotted using the recoil resolution model (Eq. 2) as shown in Figure 9. The Ti solute matrix concentration near precipitates is derived from: thermal equilibrium (solid black line), recoil resolution during single beam irradiation (solid orange line), and recoil resolution during dual beam irradiation (dashed blue line). The high temperature results will be discussed first.

The model predicts the dominance of radiation-enhanced coarsening for the single beam irradiations at elevated temperatures, most notably at 600°C. This is evidenced by the Ti concentration magnitude resulting from recoil resolution (orange line) being ~2 and 10× less than that derived from the thermal equilibrium concentration (black line) at 500°C and 600°C, respectively. Hence, though both mechanisms may be operational, coarsening is more influential.

This matches the single beam experimental results for the 500 and 600°C single beam conditions (Figure 5).

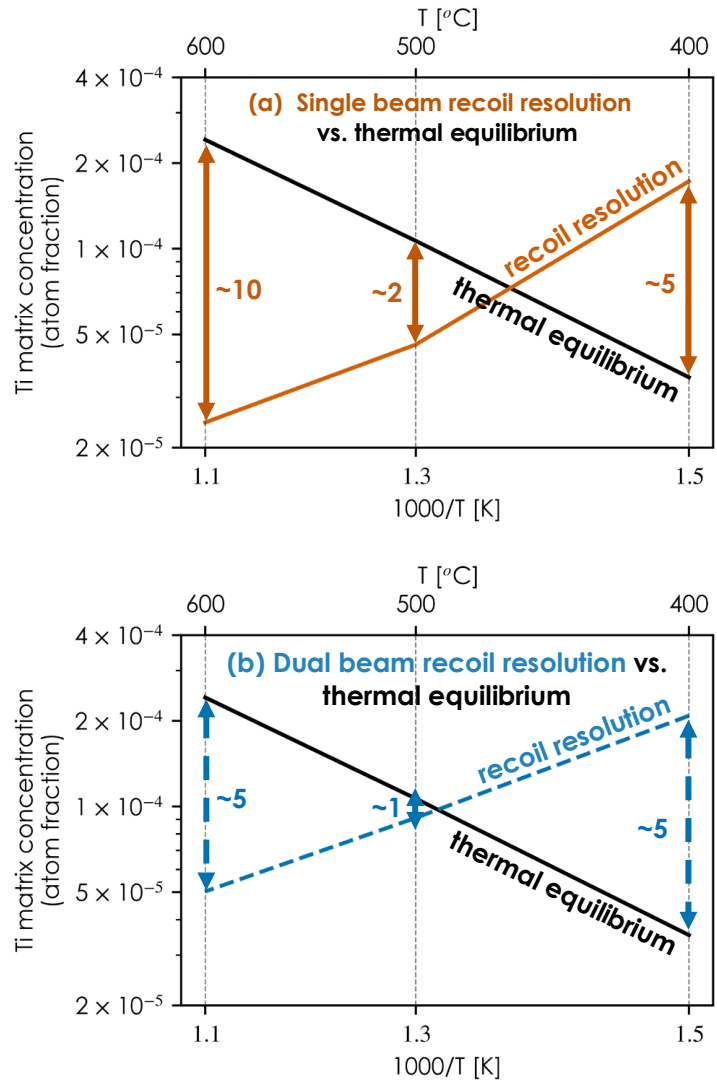

Figure 9 The Ti matrix concentration near precipitate-matrix interfaces derived from (a,b) thermal equilibrium ($c_e$, black line) and from recoil resolution under (a) single beam irradiation ($\frac{SR^2}{12D_{RED}^{Ti}}\left(1 - \frac{R}{4r_p}\right)$, orange line) and (b) dual beam irradiation ($\frac{SR^2}{12D_{RED}^{Ti}}\left(1 - \frac{R}{4r_p}\right)$, bluee line).

The model predicts the equal influence of dissolution and coarsening on precipitate stability for the 500°C dual beam condition (10 appm He/dpa) as the concentration contributions from both are equal (Figure 9b, blue line). This prediction from the model matches with the described experimental results, which showed that the size distribution and number density of MX-TiC precipitates did not change under dual beam irradiation as compared to the control condition (Figure 4). Hence, experimental and modeling results suggest that a state of stability was reached at the 500°C dual beam condition.

Similarly, the model predicts that the relative influence of recoil resolution becomes more substantial with helium co-implantation at 600°C. This is indicated by the ratio of the thermally-induced Ti concentration to the recoil resolution-induced Ti concentration decreasing from 10 to 5 for the single and dual beam irradiations at 600°C, respectively. The greater influence of recoil resolution predicted by the model with helium co-implantation bore out in the experimental results as the partial suppression of precipitate coarsening (Figure 5). However, the influence of recoil resolution was not enough to fully suppress coarsening as observed in the dual beam condition at 500°C, due to the greater diffusion of solutes at 600°C (Figure 8).

At 400°C, the same differences in magnitude were observed for the single (Figure 9a, orange line) and dual beam (Figure 9a, black line) conditions: the concentration from recoil resolution was ~5× greater than that from the thermal equilibrium concentration, solidifying that helium should have no observable impact on precipitate behavior beneath a critical temperature (i.e., solute diffusivity). Recoil resolution is the primary mechanism affecting particle stability in these conditions, though not dominant enough to cause complete dissolution as supported by the experimental results. A summary of the results is shown in Table 4.

Table 4 Dominant precipitate stability mechanisms in each irradiation condition at intermediate damage level as predicted by the Frost and Russell model and as observed experimentally.

| Irradiation Conditions | Model Predictions | Experimental Observations |
|---|---|---|
| 400°C 15 dpa - 400°C 15 dpa 10 appm He/dpa | $c_e < \frac{SR^2}{12D}\left(1 - \frac{R}{4r_p}\right)$ *Recoil resolution dominates* | - Partial dissolution of MX precipitates, no size changes<br>- Single and dual beam irradiations at 400°C displayed same precipitate behavior |
| 500°C 15 dpa 10 appm He/dpa | $c_e = \frac{SR^2}{12D}\left(1 - \frac{R}{4r_p}\right)$ *Condition of stability* | - Helium co-injection fully suppressed radiation-enhanced coarsening at 500°C such that the size distribution and statistics were equivalent to the control condition |
| 500°C 15 dpa - 600°C 15 dpa - 600°C 15 dpa 10 appm He/dpa | $c_e > \frac{SR^2}{12D}\left(1 - \frac{R}{4r_p}\right)$ *Radiation-enhanced coarsening dominates* | - Radiation-enhanced coarsening was observed in both the single beam irradiations at 500 and 600°C<br>- Helium co-injection partially suppressed radiation-enhanced coarsening at 600°C |

## 1.4 Conclusions

Using a combined experimental and modeling approach to mechanistically understand helium co-implantation effects on MX phase stability in an advanced Fe-9Cr RAFM steel alloy under fusion-relevant ion irradiation environments, the following conclusions can be drawn:

- Helium was found to significantly affect diffusion-mediated mechanisms of precipitate stability by binding to Ti solutes in the matrix and inhibiting their mobility. Specifically, helium co-injection suppressed radiation-enhanced precipitate coarsening at 500 and 600°C irradiated to 15 dpa by suppressing the matrix diffusion of Ti solutes.
- At 400°C, helium co-injection had no effect on the MX precipitate behavior because the precipitates were not in a diffusion-dominated mode of stability. As diffusion was already suppressed at this low temperature, a further suppression by the presence of helium had no discernable effect on precipitates.

- Helium co-injection was found to have no effect on the complete dissolution of MX precipitates by and past 50 dpa.

This work provides meaningful contributions to the field of phase stability under irradiation with and without helium co-addition. Importantly, the dynamic nature of precipitation under irradiation in the presence of helium is highlighted for next-generation alloys. The next paper in this series will determine the helium sequestration ability of the MX-TiC precipitates in CNA9 and the effect, if any, the sequestration had on matrix swelling.

**References**

1. Zinkle, S.J. and A. Quadling, *Extreme materials environment of the fusion "fireplace"*. MRS Bulletin, 2022. **47**(11): p. 1113-1119.
2. Ullmaier, H., *The influence of helium on the bulk properties of fusion reactor structural materials*. Nuclear Fusion, 1984. **24**(8): p. 1039-1083.
3. Was, G.S., *Fundamentals of Radiation Materials Science: Metals and Alloys*. 2nd ed. 2017: Springer.
4. Farrell, K., et al., *Modification of radiation damage microstructure by helium*. Radiation Effects, 2006. **78**(1-4): p. 277-295.
5. L.K. Mansur, W.A.C., *Mechanisms of helium interaction with radiation effects in metals and alloys: A review*. Journal of Nuclear Materials, 1983. **119**: p. 1-25.
6. Ding, J., et al., *Energetics of helium-vacancy complexes in Fe-9Cr alloys from first-principles calculations*. Journal of Nuclear Materials, 2019. **513**: p. 143-151.
7. Pitike, K.C., et al., *Helium interaction with solutes and impurities in neutron-irradiated nanostructured ferritic alloys: A first principles study*. Journal of Nuclear Materials, 2022. **566**.
8. Maziasz, P.J., *Helium trapping at Ti-rich MC particles in Ti-modified austenitic stainless steel*, ORNL, Editor. 1980.
9. Bhattacharya, A., et al., *Helium causing disappearance of a/2<111> dislocation loops in binary Fe-Cr ferritic alloys*. Journal of Nuclear Materials, 2021. **556**.
10. H. Tanigawa, E.G., T. Hirose, M. Ando, S.J. Zinkle, R. Lindau, E. Diegele, *Development of benchmark reduced activation ferritic/martensitic steels for fusion energy applications*. Nuclear Fusion, 2017. **57**(092004).
11. Sojak, S., et al., *Bubble Swelling in Ferritic/Martensitic Steels Exposed to Radiation Environment with High Production Rate of Helium*. Materials, 2021. **14**(11).

12. P. Dubuisson, D.G., J.L. Seran, *Microstructural evolution of ferritic-martensitic steels irradiated in the fast breeder reactor Phenix.* Journal of Nuclear Materials, 1993. **205**: p. 178-189.
13. Abe, F., *Precipitate design for creep strengthening of 9% Cr tempered martensitic steel for ultra-supercritical power plants.* Sci Technol Adv Mater, 2008. **9**(1): p. 013002.
14. Green, T.M.K., et al., *Effect of N2- and CO2-containing shielding gases on composition modification and carbonitride precipitation in wire arc additive manufactured grade 91 steel.* Additive Manufacturing, 2022. **56**.
15. Xiu, P., *Effects of Sink Strength and Irradiation Parameters on Defect Evolution in Additively Manufactured HT9*, in *Nuclear Engineering and Radiological Sciences)*. 2022, University of Michigan.
16. Bhattacharya, A. and S.J. Zinkle, *Cavity Swelling in Irradiated Materials*, in *Comprehensive Nuclear Materials*. 2020. p. 406-455.
17. Steven J. Zinkle, J.T.B., *Structural materials for fission & fusion energy.* Materials Today, 2009. **12**(11).
18. Zinkle, S.J., et al., *Multimodal options for materials research to advance the basis for fusion energy in the ITER era.* Nuclear Fusion, 2013. **53**(10).
19. Aitkaliyeva, A., et al., *Irradiation effects in Generation IV nuclear reactor materials*, in *Structural Materials for Generation IV Nuclear Reactors*. 2017. p. 253-283.
20. Maziasz, P.J., *Formation and stability of radiation-induced phases in neutron-irradiated austenitic and ferritic steels.* Journal of Nuclear Materials, 1989. **169**: p. 95-115.
21. E. H. Lee, A.F.R., L. K. Mansur, *Precipitation and cavity formation in austenitic stainless steels during irradiation.* Joumal of Nuclear Materials, 1981. **103 & 104**: p. 1475-1480.
22. Green, T.M.K., *MX precipitate behavior in an irradiated advanced Fe-9Cr steel: Self-ion irradiation effects on phase stability.* arXiv, 2024.
23. A.F. Rowcliffe, E.H.L., *High temperature radiation damage phenomenon in complex alloys.* Journal of Nuclear Materials, 1982. **108 & 109**: p. 306-318.
24. Mansur, L.K., *Theoretical evaluation of a mechanism of precipitate-enhanced cavity swelling during irradiation.* Philosophical Magazine A, 1981. **44**(4): p. 867-877.
25. Kesternich, W., *Helium trapping at dislocations, precipitates and grain boundaries*, in *Radiation Effects*. 1983. p. 261-273.
26. E.H. Lee, N.H.P., L.K. Mansur, *Effects of pulsed dual-ion irradiation on phase transformations and microstructure in Ti-modified austenitic alloy.* Journal of Nuclear Materials, 1983. **117**: p. 123-133.
27. T. Kimoto, H.S., *Void swelling and precipitation in a titanium-modified austenitic stainless steel under proton irradiation.* Journal of Nuclear Materials, 1985. **132**: p. 266-276.
28. Russell, K.C., *Phase stability under irradiation.* Progress in Materials Science, 1984. **28**: p. 229-434.
29. Ribis, J., *Phase Stability in Irradiated Alloys*, in *Comprehensive Nuclear Materials*. 2020. p. 265-309.
30. H.J. Frost, K.C.R., *Particle stability with recoil resolution.* Acta Metallurgica, 1982. **30**: p. 953-960.
31. *Fusion Materials Semiannual Progress Report for the Period Ending June 2021*. ORNL.
32. *Fusion Materials Semiannual Progress Report ending December 31 2021*. ORNL.

33. Mansur, L.K., et al., *Materials needs for fusion, Generation IV fission reactors and spallation neutron sources – similarities and differences.* Journal of Nuclear Materials, 2004. **329-333**: p. 166-172.
34. Was, G.S., et al., *Materials for future nuclear energy systems.* Journal of Nuclear Materials, 2019. **527**.
35. Green, T.M.K., *Effect of Damage, Temperature, and Helium on Irradiated Nanoprecipitation Stability and Helium Sequestration Ability in an Advanced Ferritic/Martensitic Fe-9Cr Steel*, https://dx.doi.org/10.7302/22241, in *Nuclear Engineering*. 2023, University of Michigan.
36. L.K. Mansur, E.H.L., P.J. Maziasz, A.P. Rowcliffe, *Control of helium effects in irradiated materials based on theory and experiment.* Journal of Nuclear Materials, 1986. **141-143**: p. 633-646.
37. Taller, S., *The Role of Damage Rate on Cavity Nucleation with Co-Injected Helium in Dual Ion Irradiated T91 Steel*, in *Nuclear Engineering*. 2020, University of Michigan.
38. Taller, S., et al., *Predicting structural material degradation in advanced nuclear reactors with ion irradiation.* Sci Rep, 2021. **11**(1): p. 2949.
39. Ziegler, J.F., M.D. Ziegler, and J.P. Biersack, *SRIM – The stopping and range of ions in matter (2010).* Nuclear Instruments and Methods in Physics Research Section B: Beam Interactions with Materials and Atoms, 2010. **268**(11-12): p. 1818-1823.
40. Stoller, R.E., et al., *On the use of SRIM for computing radiation damage exposure.* Nuclear Instruments and Methods in Physics Research Section B: Beam Interactions with Materials and Atoms, 2013. **310**: p. 75-80.
41. Taller, S., et al., *Multiple ion beam irradiation for the study of radiation damage in materials.* Nuclear Instruments and Methods in Physics Research Section B: Beam Interactions with Materials and Atoms, 2017. **412**: p. 1-10.
42. Schneider, C.A., W.S. Rasband, and K.W. Eliceiri, *NIH Image to ImageJ: 25 years of image analysis.* Nat Methods, 2012. **9**(7): p. 671-5.
43. Waskom, M., *seaborn: statistical data visualization.* Journal of Open Source Software, 2021. **6**(60).
44. Arthur Motta, D.O., *Phase Transformations Under Irradiation*, in *Light Water Reactor Materials*. 2021, American Nuclear Society.
45. Zhang, P., et al., *Interaction between helium and transition metals in vanadium: A first-principles investigation.* Nuclear Materials and Energy, 2022. **31**.

**Notes**

Portions of this supplemental are reproduced from the dissertation of T.M. Kelsy Green [35] and are provided within for clarity and completeness to the reader and to mitigate possible access issues to the dissertation document hosted and supported by the University of Michigan-Ann Arbor.

# Supplemental A: Precipitate Data

This supplemental section provides the data on the number of liftouts taken, the number of STEM-EDS maps taken, the number of precipitates counted, the number density of precipitates, the equivalent diameters of precipitates, and the volume fraction of precipitates for each single and dual beam irradiation condition in Table S.1.

Table S.1 Data on the number of liftouts taken, the number of STEM-EDS maps taken, the number of precipitates counted (N), the number density of precipitates ($\rho$), the equivalent diameters of precipitates ($d_{eq}$), and the volume fraction of precipitates (f) for each irradiation condition.

| Sample ID | # of Liftouts Taken | # of EDS Maps Taken | N | $\rho$ (m$^{-3}$) | $d_{eq}$(nm) | f |
|---|---|---|---|---|---|---|
| As-received | 1 | 5 | 90 | $(2.3\pm0.3)\times10^{21}$ | 7.8±0.3 | $(6.8\pm0.6)\times10^{-4}$ |
| 300°C, 15 dpa | 3 | 11 | 49 | $(0.9\pm0.2)\times10^{21}$ | 8.7±0.6 | $(3.9\pm1.0)\times10^{-4}$ |
| 300°C, 50 dpa | 1 | 5 | N.O. | N.O. | N.O. | N.O. |
| 400°C, 15 dpa | 3 | 9 | 67 | $(1.5\pm0.2)\times10^{21}$ | 7.6±0.4 | $(4.4\pm0.9)\times10^{-4}$ |
| 300°C, 15 dpa, 10 appm He/dpa | 2 | 9 | 56 | $(0.9\pm0.2)\times10^{21}$ | 8.4±0.4 | $(3.8\pm0.9)\times10^{-4}$ |
| 500°C, 1 dpa | 1 | 5 | 106 | $(3.3\pm0.9)\times10^{21}$ | 6.6±0.4 | $(6.9\pm0.2)\times10^{-4}$ |
| 500°C, 5 dpa | 1 | 5 | 86 | $(2.9\pm0.9)\times10^{21}$ | 7.4±0.4 | $(9.3\pm3.4)\times10^{-4}$ |
| 500°C, 15 dpa | 3 | 12 | 109 | $(1.7\pm0.3)\times10^{21}$ | 10.4±0.9 | $(11.1\pm2.3)\times10^{-4}$ |
| 500°C, 50 dpa | 2 | 9 | N.O. | N.O. | N.O. | N.O. |
| 500°C, 100 dpa | 2 | 6 | N.O. | N.O. | N.O. | N.O. |
| 500°C, 15 dpa, 10 appm He/dpa | 2 | 11 | 122 | $(1.8\pm0.1)\times10^{21}$ | 7.6±0.4 | $(5.3\pm0.5)\times10^{-4}$ |
| 500°C, 50 dpa, 10 appm He/dpa | 1 | 4 | N.O. | N.O. | N.O. | N.O. |
| 500°C, 100 dpa, 10 appm He/dpa | 2 | 4 | N.O. | N.O. | N.O. | N.O. |
| 500°C, 15 dpa, 25 appm He/dpa | 2 | 7 | 63 | $(1.5\pm0.2)\times10^{21}$ | 8.3±0.4 | $(6.1\pm1.1)\times10^{-4}$ |
| 600°C, 15 dpa | 3 | 9 | 69 | $(1.3\pm0.2)\times10^{21}$ | 14.9±2.2 | $(42.0\pm11.0)\times10^{-4}$ |
| 600°C, 15 dpa, 10 appm He/dpa | 4 | 14 | 142 | $(1.6\pm0.2)\times10^{21}$ | 11.2±1.2 | 11.2±1.2 |

**Supplemental B: Titanium and Carbon Diffusivities**

The diffusivities of Ti and C were determined to find the rate-limiting species for TiC precipitate evolution under irradiation. Table S.2 shows the calculations of the diffusivities of Ti and C as a function of temperature. The amount of Ti and C in the BCC CNA9 matrix were calculated from Thermo-Calc. Calculations were made assuming the amount of Ti and C in the matrix does not significantly alter under irradiation. Radiation-enhanced diffusion will increase the diffusion of Ti versus thermal conditions because it diffuses as a substitutional solute. However, radiation will not affect C as C is an interstitial and already diffuses quickly [46]. The diffusion of Ti was calculated using the equations outlined for single beam and dual beam irradiations in the paper. The diffusion of C in single beam irradiations was calculated using:

$$D^C_{thermal} = D^C_0 exp\left(-\frac{\Delta H_D}{k_b T}\right) \text{ [47]} \quad Eq.\ S.1$$

where $D^C_0$ is the frequency factor of diffusion of C atoms equal to $6.2 \times 10^{-7}$ m²/s and $\Delta H_D$ is the carbon diffusion activation energy equal to ~80 kJ/mol [47]. The diffusion of C in dual beam irradiations was calculated using:

$$D^C_{eff} = \frac{D^C_{thermal}}{1 + C_e exp\left(\frac{E_b^{C-He}}{kT}\right)} \quad Eq.\ S.2$$

The value of 0.33 eV (C-1NN He$_{sub}$ in BCC Fe) was chosen for $E_b^{C-He}$ because it is the most attractive binding energy for the simple C-He complex and hence the most conservative [48]. Diffusivity was calculated as the diffusion of the diffusing species (Ti or C) multiplied by its concentration in the matrix. As **Error! Reference source not found.** shows, the diffusivity values at all conditions for Ti are significantly lower than those of C. Hence, using the diffusion values for Ti instead of for C are justified.

Table S.2 This table shows the diffusivity calculations for Ti and C. Refer to the text for the equations used to calculate the values. This table justifies the use of Ti, as the rate-limiting species, for diffusion calculations regarding precipitate stability in the proceeding sections.

| Temperature (°C) | C concentration in the matrix (atom fraction) | Ti concentration in the matrix (atom fraction) | $D_C$ in the matrix ($\times 10^{-7}$ m$^2$/s) | | $D_{Ti}$ in the matrix ($\times 10^{-19}$ m$^2$/s) | | Diffusivity of C in the matrix (m$^2$/s) | | Diffusivity of Ti in the matrix ($\times 10^{-22}$ m$^2$/s) | |
|---|---|---|---|---|---|---|---|---|---|---|
| | | | SI ($D_{thermal}^{C}$) | DI ($D_{eff}^{C}$) | SI ($D_{RED}^{Ti}$) | DI ($D_{RED,He}^{Ti}$) | SI | DI | SI | DI |
| 300 | 3×10$^{-12}$ | 9.14×10$^{-6}$ | 6.1 | - | 5.1±0.009 | - | 1.8×10$^{-18}$ | - | 0.05±0.00009 | - |
| 400 | 1.21×10$^{-9}$ | 3.53×10$^{-5}$ | 6.1 | 5.9 | 8.8±0.02 | 0.6±0.02 | 7.4×10$^{-16}$ | 7.1×10$^{-16}$ | 0.3±0.001 | 0.002±0.00008 |
| 500 | 9.75×10$^{-8}$ | 1.06×10$^{-4}$ | 6.1 | 6.0 | 13.4±0.1 | 1.4±0.04 | 6.0×10$^{-14}$ | 5.8×10$^{-14}$ | 1.4±0.01 | 0.2±0.004 |
| 600 | 2.22×10$^{-6}$ | 2.41×10$^{-4}$ | 6.1 | 6.1 | 12.3±0.4 | 2.6±0.1 | 1.4×10$^{-12}$ | 1.3×10$^{-12}$ | 3.0±0.09 | 0.6±0.02 |

## Supplemental C: Calculation of Sink Strength

The sink strengths for interstitials and vacancies, $k^2_{(i,v)}$, for interstitials and vacancies are the sum of the sink strengths of the individually measured sinks, accounting for the bias of interstitials for dislocations:

$$k^2_v = \sum_s k^2_{vs} = k^2_{gb} + k^2_{cav} + k^2_{ppts} + k^2_{dis} \qquad Eq.\ S.3$$

$$k^2_i = \sum_s k^2_{is} = k^2_{gb} + k^2_{cav} + k^2_{ppts} + k^2_{dis} \qquad Eq.\ S.4$$

$$k^2 = k^2_v + k^2_i \qquad Eq.\ S.5$$

where $k^2_{gb}$ is the grain boundary sink strength from PAGs and laths, $k^2_{cav}$ is the sink strength from voids and bubbles, $k^2_{ppts}$ is the precipitate sink strength from the MX-TiC precipitates, and $k^2_{dis}$ is the dislocation sink strength.

The sink strength of cavities was calculated by assuming all cavities below 5 nm are helium-pressurized bubbles and all cavities over 5 nm are voids:

$$k^2_{cav} = k^2_{bub} + k^2_{void} \qquad Eq.\ S.6$$

$$k^2_{bub} = k^2_{bub,i} + k^2_{bub,v} \qquad Eq.\ S.7$$

$$k^2_{bub,i} = 4Y_i \pi \rho_{bub} r_{bub} \qquad Eq.\ S.8$$

$$k^2_{bub,v} = 4Y_v \pi \rho_{bub} r_{bub} \qquad Eq.\ S.9$$

$$Y_i = \frac{(\alpha_{v,cavity} - \alpha_{i,cavity})(T_m/T)^{1/3}}{r_v + r_{bub} + \alpha_{bub,v}(T_m/T)^{1/3}} \qquad Eq.\ S.10$$

$$Y_v = 1 \qquad Eq.\ S.11$$

$$k^2_{void} = 4\pi \rho_{void} r_{void} \qquad Eq.\ S.12$$

where $k_{bub,i}^2$ is the sink strength of bubbles for interstitials, $k_{bub,v}^2$ is the sink strength of bubbles for vacancies, $k_{bub}^2$ is the total bubble sink strength, $\rho_{bub}$ is the number density of bubbles, $r_{bub}$ is the average radius of bubbles, $Y_i$ is the bias of bubbles for interstitials, $Y_v$ is the bias of bubbles for vacancies, $r_v$ is the radius of a vacancy (taken as the radius of an Fe atom), and $\alpha_{v,cavity}$ and $\alpha_{i,cavity}$ are fitting parameters derived from Ref. [49]. $\alpha_{v,cavity}$ and $\alpha_{i,cavity}$ are equal to 0.83Å and 3.19 Å, respectively. Voids are assumed to be neutral sinks.

The sink strength of MX-TiC precipitates was calculated as [50]:

$$k_{ppt}^2 = 4\pi\rho_{ppt}r_{ppt} \qquad Eq.\ S.13$$

where $\rho_{ppt}$ is the number density of precipitates and $r_{ppt}$ is the average equivalent radius of the precipitates [3].

The sink strength of dislocations was calculated as:

$$k_{dis}^2 = k_{loop}^2 + k_{line}^2 \qquad Eq.\ S.14$$

$$k_{(loop,line),i}^2 = 2B_i\pi\rho_{(loop,line)}r_{(loop,line)} \qquad Eq.\ S.15$$

$$k_{(loop,line),v}^2 = 2B_v\pi\rho_{(loop,line)}r_{(loop,line)} \qquad Eq.\ S.16$$

$$k_{(loop,line)}^2 = k_{(loop,line),i}^2 + k_{(loop,line),v}^2 \qquad Eq.\ S.17$$

where $k_{loop}^2$ is the sink strength of dislocation loops, $k_{line}^2$ is the sink strength of dislocation lines, $B_i$ is the bias factor of dislocations for interstitials, and $B_v$ is the bias factor of dislocations for vacancies. $B_v$ is assumed to be 1, meaning dislocations are not biased toward vacancies, based off of literature [3]. $B_i$ was calculated to be 8% from Ref. [49]. $B_i$ was calculated to satisfy the bias-driven criterion at 500°C. The bias-driven criterion in Ref. [49] is defined as a model of cavity behavior in which growth of small cavities is driven by helium accumulation until a critical radius is reached, whereby cavities grow via bias-driven partitioning of point defects. The critical bubble radius is the radius of cavities where the cavities transition from being stabilized by helium gas to a vacancy-biased growth regime. The critical bubble radius ($r_{crit}^{bias-driven}$) was determined experimentally by examining the cavity size distribution at 500°C to 15 dpa with 10 appm He/dpa in Figure S.1. $r_{crit}^{bias-driven}$ was input into Eq. S.18, where the biases for dislocations and cavities are set equal. $B_i$ is solved for to allow this criterion to be true:

$$r_{crit}^{bias-driven} = \frac{(\alpha_{i,cavity} - \alpha_{v,cavity})(T_m/T)^{1/3} - B_i(r_{vac} + \alpha_{v,cavity}(T_m/T)^{1/3})}{B_i} \quad Eq.\ S.18$$

It was determined that the critical bubble radius from this equation matched experimental data at 500°C if $B_i$ was equal to 8%. It was determined this bias did not change as a function of irradiation parameters.

The sink strength of grain boundaries was calculated using the following equation:

$$k_{gb}^2 = k_{PAG}^2 + k_{lath}^2 \quad Eq.\ S.19$$

where

$$k_{PAG,lath}^2 = \frac{6\sqrt{k_{inside\ grain}^2}}{d_{PAG,lath}} \quad Eq.\ S.20$$

$$k_{inside\ grain}^2 = k_{cav}^2 + k_{ppts}^2 + k_{dis}^2 \text{ (SI)} \quad Eq.\ S.21$$

$$k_{inside\ grain}^2 = k_{cav}^2 + k_{ppts}^2 + k_{dis}^2 + k_{bub}^2 + k_{void}^2 \text{ (DI)} \quad Eq.\ S.22$$

All error associated with the input values to the equations above were propagated with the use of the uncertainties Python package. The results of the calculations for point defect concentrations are shown in Table S.3. Sink strength calculations for select single ion irradiation experiments are shown in Table S.4 and for dual ion irradiation experiments in Table S.5.

The measured values of precipitates from this work were used in Eq. S10. Dislocation loop size and density for 300°C was taken from Ref. [51]. Ref. [51] irradiated Grade 91 to 30 dpa at 300°C with a dose rate $2\times10^{-3}$ dpa/s. As dislocation loop size and density are a function of temperature, dose, and dose rate, it can be assumed that the value used here will not exactly match CNA9's value at 300°C but were appropriate for use due to lack of literature data [52]. Values for dislocation loop size and density and for dislocation line density for all temperatures besides 300°C were taken from Ref. [37]. Dislocation line density for 300°C assumed to be same as at 400°C, as data was lacking from literature. Values from Ref. [37] did not match the experimental conditions in this work exactly. Ref. [37] used Grade 91 irradiated at 406°C-16.6dpa-$7\times10^{-4}$ dpa/s-4.3 appm He/dpa, 480°C-16.6dap-$7\times10^{-4}$ dpa/s-4.3 appm He/dpa, and 570°C-15.4dpa-$7\times10^{-4}$ dpa/s-4.3 appm He/dpa. Hence, values from Ref. [37] input for 400°C in these calculations were taken from the experiment run at 406°C. Values input for 500°C in these calculations were taken from the experiment run at 480°C. Values input for 600°C in these calculations were taken from the experiment run at 570°C. Errors for dislocation loop size and density from Ref. [37] were reported in the reference. Grade 91 is an appropriate surrogate material due to the similarity in composition to CNA9 and the similar grain and lattice structures. Hence, it is assumed that values of dislocation line densities of Grade 91 can be used for CNA9, within appropriate reason. However, the helium rate used in this thesis was 2.3× greater than used in Ref. [5]. Helium implantation level has been shown to affect the dislocation structure by altering the ratio of dislocation loop type in BCC Fe-Cr steel alloys, but not by altering the number density of total dislocation loops [11]. Hence, it is assumed that using the dislocation loop size and densities from Ref. [5] is appropriate within reason. Values for the prior austenite grain (PAG) size was taken from Ref. [37] and the martensite lath size was taken from Ref. [53]. Error for the lath size was assumed to be 10% to cover the spread of lath sizes found in literature [14].

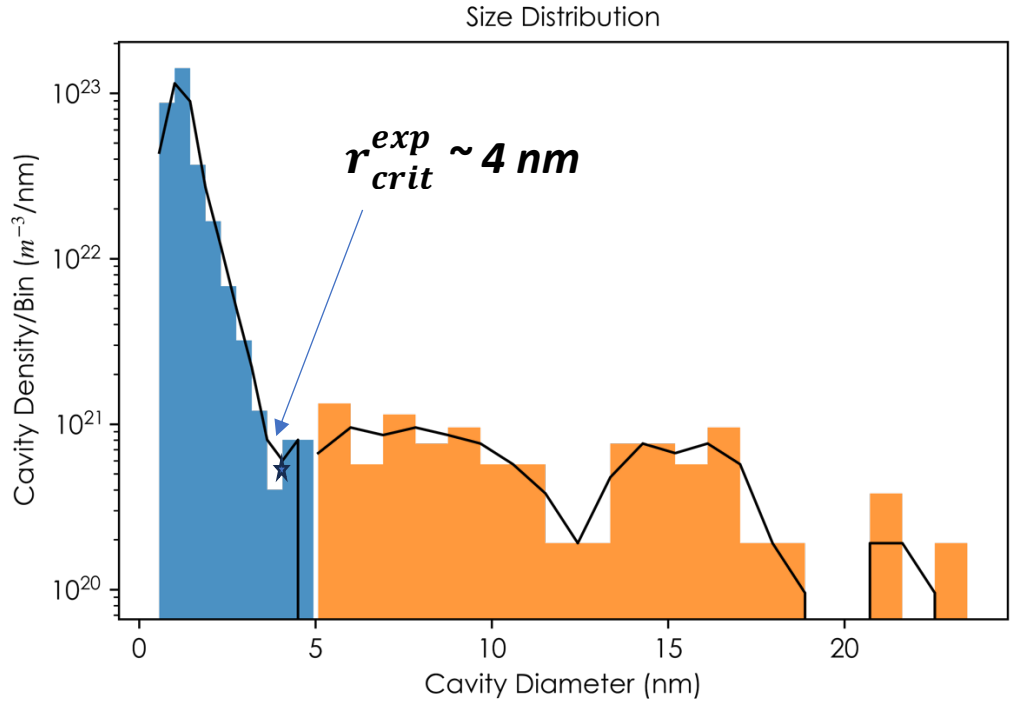

Figure S.1. The size distribution of cavities for the dual beam condition at 500°C and 15 dpa with 10 appm He/dpa.

Table S.4 Values input to calculate sink strength for single ion irradiation conditions in the temperature series to 15 dpa with $7\times10^{-4}$ dpa/s. N.C. means not calculated.

| | | | | |
|---|---|---|---|---|
| Dislocation Lines | Number density (m$^{-3}$) | 300°C: $3.8\times10^{14}$<br>400°C: $3.8\times10^{14}$<br>500°C: $0.4\times10^{14}$<br>600°C: N.O. | $k^2_{line}$ (m$^{-2}$) | 300°C: $7.9\times10^{14}$<br>400°C: $7.9\times10^{14}$<br>500°C: $8.7\times10^{13}$<br>600°C: N.C. |
| Dislocation Loops | Number density (m$^{-3}$) | 300°C: $4.1\times10^{22}$<br>400°C: $12\times10^{21}$<br>500°C: $0.46\times10^{22}$<br>600°C: N.O. | $k^2_{loop}$ (m$^{-2}$) | 300°C: $(2.3\pm1.1)\times10^{15}$<br>400°C: $(1.6\pm0.2)\times10^{15}$<br>500°C: $(1.7\pm0.2)\times10^{14}$<br>600°C: N.C. |
| | Radius (nm) | 300°C: $4.3\pm2.1$<br>400°C: $10.1\pm1.2$<br>500°C: $29\pm3.8$<br>600°C: N.O. | | |
| MX-TiC precipitates | Number density (m$^{-3}$) | 300°C: $(8.8\pm1.8)\times10^{20}$<br>400°C: $(1.4\pm0.2)\times10^{21}$<br>500°C: $(1.7\pm0.3)\times10^{21}$<br>600°C: $(1.3\pm0.2)\times10^{21}$ | $k^2_{MX}$ (m$^{-2}$) | 300°C: $(4.8\pm1.0)\times10^{13}$<br>400°C: $(6.9\pm1.0)\times10^{13}$<br>500°C: $(1.1\pm0.2)\times10^{14}$<br>600°C: $(1.2\pm0.2)\times10^{14}$ |
| | Radius (nm) | 300°C: $4.4\pm0.09$<br>400°C: $3.8\pm0.06$<br>500°C: $5.2\pm0.1$<br>600°C: $7.4\pm0.4$ | | |
| Grain Boundaries | Diameter of PAGs (μm) | 15 | $k^2_{GB}$ (m$^{-2}$) | 300°C: $(7.0\pm0.7)\times10^{14}$<br>400°C: $(6.1\pm0.6)\times10^{14}$<br>500°C: $(2.4\pm0.2)\times10^{14}$<br>600°C: $(1.4\pm0.1)\times10^{14}$ |
| | Diameter of laths (nm) | $500\pm50$ | | |
| Total | | | $k^2_{tot}$ (m$^{-2}$) | 300°C: $(4.6\pm1.1)\times10^{15}$<br>400°C: $(3.7\pm0.2)\times10^{15}$<br>500°C: $(9.7\pm0.6)\times10^{14}$<br>600°C: $(5.1\pm0.5)\times10^{14}$ |

Table S.5 Values input to calculate sink strength for dual ion irradiation conditions in the temperature series to 15 dpa with $7\times10^{-4}$ dpa/s and 10 appm He/dpa. N.C. means not calculated.

| | | | | |
|---|---|---|---|---|
| Dislocation Lines | Number density (m$^{-3}$) | 400°C: $3.8\times10^{14}$<br>500°C: $0.4\times10^{14}$<br>600°C: N.O. | $k^2_{line}$ (m$^{-2}$) | 400°C: $7.9\times10^{14}$<br>500°C: $8.7\times10^{13}$<br>600°C: N.C. |
| Dislocation Loops | Number density (m$^{-3}$) | 400°C: $12\times10^{21}$<br>500°C: $0.46\times10^{22}$<br>600°C: N.O. | $k^2_{loop}$ (m$^{-2}$) | 400°C: $(1.6\pm0.2)\times10^{15}$<br>500°C: $(1.7\pm0.2)\times10^{14}$<br>600°C: N.C. |
| | Radius (nm) | 400°C: $10.1\pm1.2$<br>500°C: $29\pm3.8$<br>600°C: N.O. | | |
| MX-TiC precipitates | Number density (m$^{-3}$) | 400°C: $(9.2\pm1.8)\times10^{20}$<br>500°C: $(1.8\pm0.1)\times10^{21}$<br>600°C: $(1.6\pm0.2)\times10^{21}$ | $k^2_{MX}$ (m$^{-2}$) | 400°C: $(4.9\pm0.9)\times10^{13}$<br>500°C: $(8.5\pm0.7)\times10^{13}$<br>600°C: $(1.1\pm0.2)\times10^{14}$ |
| | Radius (nm) | 400°C: $4.2\pm0.07$<br>500°C: $3.8\pm0.06$<br>600°C: $5.6\pm0.2$ | | |
| Grain Boundaries | Diameter of PAGs (μm) | 15 | $k^2_{GB}$ (m$^{-2}$) | 400°C: $(6.6\pm0.6)\times10^{14}$<br>500°C: $(3.7\pm0.4)\times10^{14}$<br>600°C: $(2.4\pm0.2)\times10^{14}$ |
| | Diameter of laths (nm) | $500\pm50$ | | |
| Bubbles | Number density (m$^{-3}$) | 400°C: $(3.3\pm0.2)\times10^{22}$<br>500°C: $(1.8\pm0.07)\times10^{22}$<br>600°C: $(1.9\pm0.07)\times10^{22}$ | $k^2_{bub}$ (m$^{-2}$) | 400°C: $(4.8\pm0.3)\times10^{14}$<br>500°C: $(5.5\pm0.2)\times10^{14}$<br>600°C: $(3.6\pm0.1)\times10^{14}$ |
| | Radius (nm) | 400°C: $0.47\pm0.0049$<br>500°C: $0.99\pm0.0039$<br>600°C: $0.65\pm0.0040$ | | |
| Voids | Number density (m$^{-3}$) | 400°C: N.O.<br>500°C: $(1.4\pm0.05)\times10^{21}$<br>600°C: N.O. | $k^2_{void}$ (m$^{-2}$) | 400°C: N.O.<br>500°C: $(1.0\pm0.0)\times10^{14}$<br>600°C: N.O. |
| | Radius (nm) | 400°C: N.O.<br>500°C: $5.70\pm0.0019$<br>600°C: N.O. | | |
| Total | | | $k^2_{tot}$ (m$^{-2}$) | 400°C: $(4.3\pm0.2)\times10^{15}$<br>500°C: $(1.9\pm0.1)\times10^{15}$<br>600°C: $(1.1\pm0.1)\times10^{15}$ |

# References

1. Zinkle, S.J. and A. Quadling, *Extreme materials environment of the fusion "fireplace"*. MRS Bulletin, 2022. **47**(11): p. 1113-1119.
2. Ullmaier, H., *The influence of helium on the bulk properties of fusion reactor structural materials*. Nuclear Fusion, 1984. **24**(8): p. 1039-1083.
3. Was, G.S., *Fundamentals of Radiation Materials Science: Metals and Alloys*. 2nd ed. 2017: Springer.
4. Farrell, K., et al., *Modification of radiation damage microstructure by helium*. Radiation Effects, 2006. **78**(1-4): p. 277-295.
5. L.K. Mansur, W.A.C., *Mechanisms of helium interaction with radiation effects in metals and alloys: A review*. Journal of Nuclear Materials, 1983. **119**: p. 1-25.
6. Ding, J., et al., *Energetics of helium-vacancy complexes in Fe-9Cr alloys from first-principles calculations*. Journal of Nuclear Materials, 2019. **513**: p. 143-151.
7. Pitike, K.C., et al., *Helium interaction with solutes and impurities in neutron-irradiated nanostructured ferritic alloys: A first principles study*. Journal of Nuclear Materials, 2022. **566**.
8. Maziasz, P.J., *Helium trapping at Ti-rich MC particles in Ti-modified austenitic stainless steel*, ORNL, Editor. 1980.
9. Bhattacharya, A., et al., *Helium causing disappearance of a/2<111> dislocation loops in binary Fe-Cr ferritic alloys*. Journal of Nuclear Materials, 2021. **556**.
10. H. Tanigawa, E.G., T. Hirose, M. Ando, S.J. Zinkle, R. Lindau, E. Diegele, *Development of benchmark reduced activation ferritic/martensitic steels for fusion energy applications*. Nuclear Fusion, 2017. **57**(092004).
11. Sojak, S., et al., *Bubble Swelling in Ferritic/Martensitic Steels Exposed to Radiation Environment with High Production Rate of Helium*. Materials, 2021. **14**(11).
12. P. Dubuisson, D.G., J.L. Seran, *Microstructural evolution of ferritic-martensitic steels irradiated in the fast breeder reactor Phenix*. Journal of Nuclear Materials, 1993. **205**: p. 178-189.
13. Abe, F., *Precipitate design for creep strengthening of 9% Cr tempered martensitic steel for ultra-supercritical power plants*. Sci Technol Adv Mater, 2008. **9**(1): p. 013002.
14. Green, T.M.K., et al., *Effect of N2- and CO2-containing shielding gases on composition modification and carbonitride precipitation in wire arc additive manufactured grade 91 steel*. Additive Manufacturing, 2022. **56**.
15. Xiu, P., *Effects of Sink Strength and Irradiation Parameters on Defect Evolution in Additively Manufactured HT9*, in *Nuclear Engineering and Radiological Sciences)*. 2022, University of Michigan.
16. Bhattacharya, A. and S.J. Zinkle, *Cavity Swelling in Irradiated Materials*, in *Comprehensive Nuclear Materials*. 2020. p. 406-455.
17. Steven J. Zinkle, J.T.B., *Structural materials for fission & fusion energy*. Materials Today, 2009. **12**(11).
18. Zinkle, S.J., et al., *Multimodal options for materials research to advance the basis for fusion energy in the ITER era*. Nuclear Fusion, 2013. **53**(10).
19. Aitkaliyeva, A., et al., *Irradiation effects in Generation IV nuclear reactor materials*, in *Structural Materials for Generation IV Nuclear Reactors*. 2017. p. 253-283.

20. Maziasz, P.J., *Formation and stability of radiation-induced phases in neutron-irradiated austenitic and ferritic steels.* Journal of Nuclear Materials, 1989. **169**: p. 95-115.
21. E. H. Lee, A.F.R., L. K. Mansur, *Precipitation and cavity formation in austenitic stainless steels during irradiation.* Joumal of Nuclear Materials, 1981. **103 & 104**: p. 1475-1480.
22. Green, T.M.K., *MX precipitate behavior in an irradiated advanced Fe-9Cr steel: Self-ion irradiation effects on phase stability.* arXiv, 2024.
23. A.F. Rowcliffe, E.H.L., *High temperature radiation damage phenomenon in complex alloys.* Journal of Nuclear Materials, 1982. **108 & 109**: p. 306-318.
24. Mansur, L.K., *Theoretical evaluation of a mechanism of precipitate-enhanced cavity swelling during irradiation.* Philosophical Magazine A, 1981. **44**(4): p. 867-877.
25. Kesternich, W., *Helium trapping at dislocations, precipitates and grain boundaries*, in *Radiation Effects*. 1983. p. 261-273.
26. E.H. Lee, N.H.P., L.K. Mansur, *Effects of pulsed dual-ion irradiation on phase transformations and microstructure in Ti-modified austenitic alloy.* Journal of Nuclear Materials, 1983. **117**: p. 123-133.
27. T. Kimoto, H.S., *Void swelling and precipitation in a titanium-modified austenitic stainless steel under proton irradiation.* Journal of Nuclear Materials, 1985. **132**: p. 266-276.
28. Russell, K.C., *Phase stability under irradiation.* Progress in Materials Science, 1984. **28**: p. 229-434.
29. Ribis, J., *Phase Stability in Irradiated Alloys*, in *Comprehensive Nuclear Materials*. 2020. p. 265-309.
30. H.J. Frost, K.C.R., *Particle stability with recoil resolution.* Acta Metallurgica, 1982. **30**: p. 953-960.
31. *Fusion Materials Semiannual Progress Report for the Period Ending June 2021*. ORNL.
32. *Fusion Materials Semiannual Progress Report ending December 31 2021*. ORNL.
33. Mansur, L.K., et al., *Materials needs for fusion, Generation IV fission reactors and spallation neutron sources – similarities and differences.* Journal of Nuclear Materials, 2004. **329-333**: p. 166-172.
34. Was, G.S., et al., *Materials for future nuclear energy systems.* Journal of Nuclear Materials, 2019. **527**.
35. Green, T.M.K., *Effect of Damage, Temperature, and Helium on Irradiated Nanoprecipitation Stability and Helium Sequestration Ability in an Advanced Ferritic/Martensitic Fe-9Cr Steel,* https://dx.doi.org/10.7302/22241, in *Nuclear Engineering*. 2023, University of Michigan.
36. L.K. Mansur, E.H.L., P.J. Maziasz, A.P. Rowcliffe, *Control of helium effects in irradiated materials based on theory and experiment.* Journal of Nuclear Materials, 1986. **141-143**: p. 633-646.
37. Taller, S., *The Role of Damage Rate on Cavity Nucleation with Co-Injected Helium in Dual Ion Irradiated T91 Steel*, in *Nuclear Engineering*. 2020, University of Michigan.
38. Taller, S., et al., *Predicting structural material degradation in advanced nuclear reactors with ion irradiation.* Sci Rep, 2021. **11**(1): p. 2949.
39. Ziegler, J.F., M.D. Ziegler, and J.P. Biersack, *SRIM – The stopping and range of ions in matter (2010).* Nuclear Instruments and Methods in Physics Research Section B: Beam Interactions with Materials and Atoms, 2010. **268**(11-12): p. 1818-1823.

40. Stoller, R.E., et al., *On the use of SRIM for computing radiation damage exposure.* Nuclear Instruments and Methods in Physics Research Section B: Beam Interactions with Materials and Atoms, 2013. **310**: p. 75-80.
41. Taller, S., et al., *Multiple ion beam irradiation for the study of radiation damage in materials.* Nuclear Instruments and Methods in Physics Research Section B: Beam Interactions with Materials and Atoms, 2017. **412**: p. 1-10.
42. Schneider, C.A., W.S. Rasband, and K.W. Eliceiri, *NIH Image to ImageJ: 25 years of image analysis.* Nat Methods, 2012. **9**(7): p. 671-5.
43. Waskom, M., *seaborn: statistical data visualization.* Journal of Open Source Software, 2021. **6**(60).
44. Arthur Motta, D.O., *Phase Transformations Under Irradiation*, in *Light Water Reactor Materials*. 2021, American Nuclear Society.
45. Zhang, P., et al., *Interaction between helium and transition metals in vanadium: A first-principles investigation.* Nuclear Materials and Energy, 2022. **31**.
46. H. J. Frost, K.C.R., *Recoil resolution and particle stability under irradiation.* Joumal of Nuclear Materials, 1981. **103 & 104**: p. 1427-143.
47. A. A. Vasilyev, P.A.G., *Carbon diffusion coefficient in alloyed ferrite.* Materials Physics and Mechanics, 2018. **39**: p. 111-119.
48. Zhang, Y., et al., *Effect of carbon and alloying solute atoms on helium behaviors in α -Fe.* Journal of Nuclear Materials, 2017. **484**: p. 103-109.
49. Kohnert, A.A., M.A. Cusentino, and B.D. Wirth, *Molecular statics calculations of the biases and point defect capture volumes of small cavities.* Journal of Nuclear Materials, 2018. **499**: p. 480-489.
50. A. D. Brailsford, L.K.M., *The effect of precipitate-matrix interface sinks on the growth of voids in the matrix.* Journal of Nuclear Materials, 1981. **103 & 104**: p. 1403-1408.
51. Duan, J., et al., *Effect of grain size on the irradiation response of Grade 91 steel subjected to Fe ion irradiation at 300°C.* Journal of Materials Science, 2022. **57**(28): p. 13767-13778.
52. Chen, Y., *Irradiation Effects of HT-9 Martensitic Steel.* Nuclear Engineering and Technology, 2013. **45**(3): p. 311-322.
53. Klueh, R.L., *Elevated-Temperature Ferritic and Martensitic Steels and Their Application to Future Nuclear Reactors*. 2004, ORNL.